# Title: A computational insight of the improved nicotine binding with ACE2-SARS-CoV-2 complex with its clinical impact

Selvaa Kumar C[1*], Senthil Arun Kumar[2], Haiyan Wei[2]

[1]School of Biotechnology and Bioinformatics, D. Y. Patil Deemed to be University, Sector-15, CBD Belapur. Navi Mumbai-400614, India.

[2]Department of Endocrinology and Metabolism, Genetics, Henan children's hospital (Children's hospital affiliated to Zhengzhou University), No-33, Longhu Waihuan East road, Zhengzhou-450018, China.

*Correspondence to:
Dr.Selvaa Kumar C
selvaakumar.c@dypatil.edu

**Abstract**

Smokers being witnessed with the mild adverse clinical symptoms of SARS-CoV-2, the in-silico study is intended to explore the effect of nicotine binding to the soluble angiotensin converting enzyme II (ACE2) receptor with or without SARS-CoV-2 binding. Nicotine established a stable interaction with the conserved amino acid residues: Asp382, Gly405, His378 and Tyr385 through His401 of the soluble ACE2 that seals its interaction with the INS1. Also, nicotine binding has significantly reduced the affinity score of ACE2 with INS1 to -12.6 kcal/mol (*versus* -15.7 kcal/mol without nicotine) and the interface area to 1933.6 Å$^2$ (*versus* 2057.3Å$^2$ without nicotine). Nicotine exhibited a higher binding affinity score with ACE2-SARS-CoV-2 complex with -6.33 kcal/mol (Vs -5.24 kcal/mol without SARS-CoV-2) and a lowered inhibitory contant value of 22.95 µM (Vs 151.69 µM without SARS-CoV). Eventhough ACE2 is not a potential receptor for nicotine binding in the healthy people, in COVID19 patients, it may exhibit better binding affinity with the ACE2 receptor. In overall, nicotine's strong preference for ACE2-SARS-CoV-2 complex might drastically reduce the SARS-CoV-2 virulence by intervening the ACE2 conserved residues interaction with the spike (S1) protein of SARS-CoV-2.

**Keywords:** Nicotine; ACE2; SARS-CoV-2; ACE2-SARS-CoV-2 complex; smokers.

**Introduction**

COVID-19, a highly contagious virus strain of SARS-CoV family of Wuhan origin has become a major life-threat to the global population (1). Clinical studies have been widely conducted using various anti-viral agents; viral-protein specific monoclonal antibodies; indigenous therapeutic bio-actives; and convalescent plasma transfusion to attenuate the human to human transmission chain of COVID-19 and its virulence across the global countries (2, 3). Of all the clinical-therapeutic targets, including the direct COVID-19 based targets (4), the host-based target of angiotensin-converting enzyme 2 (ACE2) receptor has driven major attention among the clinical researchers for the COVID-19 management (5, 6). This is because of ACE2 acting as a potent agonist for COVID-19 paving its initial entry and the colonization in the upper respiratory system (6). ACE2 is widely expressed in the arterial-venous endothelial cells, including the alveolar epithelial cells of lungs, enterocytes of the small intestine and arterial smooth muscle cells facilitate the conversion of angiotensin-converting enzyme II to I (7, 8). It has been proposed that COVID-19 binding with ACE2 would affect its stability, as well as its biological function, resulted in increased angiotensin-II levels causing severe respiratory ailments on the COVID-19 patients (8, 9). Also, the increased ACE-2 expression mediated by the COVID-19 entry would contribute to the respiratory dysfunction by agitating the immune response with the increased pro-inflammatory cytokines production, especially IL-10, IL-8 and IL-1β (6, 10). With the ACE2 being highly expressed in lungs among the smokers than the healthy individuals (11), clinicians from China have found the smoking habits doesn't influence the mortality rate among the smokers over the non-smokers (12). This clinical witness has enlightened nicotine with a bitter taste to benefit the smokers with the likely chances of reducing the COVID-19 virulence (12). Since the whole universe is seeking for a proficient clinical agent to tackle the emerging mortality of COVID-19 patients across the global countries, we have performed this in-silico study to evaluate the effect of nicotine on the soluble ACE2 receptor sensitized with the COVID-19 agonist. We find this in-silico study to be exciting as well as crucial because of its significance; that will unveil the clinical understanding of nicotine binding intervening the soluble ACE2 interactions with the crucial spike (S1) membrane protein of COVID-19. This computational understanding would indeed help the clinicians to prefer nicotine as a supplementary bio-active to treat or prevent the deteriorating health conditions of COVID-19 patients, especially the non-smokers, who are susceptible to detrimental clinical symptoms preceded by the unprecedented pro-inflammatory cytokines production.

**Materials and Methods**

**Homology modelling and structural characterization of human ACE2 and spike (S1) protein of SARS-COV-2 (of India origin).**

Angiotensin-converting enzyme 2 (ACE2) protein sequence of *Homo sapiens* was obtained from the Uniprot database (Accession Number: Q9BYF1) (13). ACE2 is available in complete (aa:18-805) and soluble form (aa:18-708). They have three distinct domains viz. extracellular domain (aa 18-740); helical domain (aa 741-761) and a cytoplasmic domain (aa 792-805). Most importantly, the interface regions of ACE2 ( aa 30-41; aa 82-84 and aa 353-357) were preferred by S1 protein of SARS-COV-2 for protein-protein interaction. Search for a potential template for ACE2 from Protein Data Bank (14) listed PDB ID: 6m18 (15) with many missing residues (14, 15). As a corrective measure, homology modelling was initiated using SWISS-MODEL online server to build the complete 3D structure of ACE2 protein (16). From the modelled ACE2 protein, the soluble ACE2 (sACE2) (aa region: 18-708) was derived using CHIMERA software for further analysis. Indian SARS-COV-2 sequence (INS1) spike protein (MT012098) (17) was downloaded from NCBI and further considered for homology modelling using SWISS-MODEL online server. Modelled ACE2 and INS1 3D structures were energy minimized using CHIMERA software (18). Finally, modelled ACE2 and INS1 were chosen for structure validation using PROCHECK-Ramachandran plot server (19).

**Protein-Protein docking of soluble ACE2 with the spike (S1) membrane protein of SARS-CoV-2**

sACE2 protein and INS1 were preferred for protein-protein docking using HADDOCK server (20). As per the Uniprot report, the recruited interface region was within the range of aa 30-41, aa 82-84 and aa 353-357 for protein-protein docking. For INS1 protein, the residues engaged with receptor interaction were scrutinized by the literature study (21, 22) wherein the entire Receptor Binding Domain (aa 319-541) was chosen for the docking study. In the context of recruting passive residues, the checkbox was selected for ACE2 and INS1. Thus residues proximal to the active loci would be marked as passive residues over here. In-total about ten clusters of 4 poses each were generated by the HADDOCK online server. Among which the cluster with the least HADDOCK score was selected for the binding energy analysis using PDBePISA (23). This tools helps in calculating the protein-protein interface and the overall binding affinity between the two proteins after docking. Docked poses and hydrogen bonds were examined using the Discovery Studio Visualizer (24).

**Docking of nicotine with the homology modelled ACE2-SARS-CoV-2 complex**

Modelled ACE2 was docked with nicotine downloaded from Protein Data Bank (PDB ID:1UW6) (25). Here nicotine was docked with ACE2 receptor based on the study of Joshua et al. wherein they have clearly mentioned that nicotine may break the interaction between ACE2 and SARS-COV2 (26). AutoDock Tools of version 1.5.6 was used for protein-ligand docking (27) (Morris, G et al, 2009). Literature evidence has shown the active site residues for ACE2 to be Arg273, His345, Pro346, Glu375, His505 and Tyr515 (21). Here we added Kollman and Gasteiger charges to the protein and ligands, respectively. The grid box was placed within the active site residues which are six in number. The size of the grid box was 44 Å, 116 Å and 48Å for x, y and z, respectively. Furthermore, the grid centre was customized to -0.861, -6.417 and 3.778 for x, y and z, respectively. AutoGrid 4.0 and AutoDock 4.0 programs were used to generate the grid maps. The best ten conformers were generated using a Lamarckian Genetic Algorithm. Following, the sACE2-SARS-COV-2 complex was scrutinized for protein-ligand docking for which the grid box was generated within the sACE2 site. The generated grid box was localized within the active site of a box size of 72Å, 44 Å and 82Å for x, y and z, respectively. Furthermore, the grid centre was customized to 42.694, -9.11 and 1.389 for x, y and z, respectively. Meanwhile, the AutoGrid 4.0 and AutoDock 4.0 programs were employed to generate the grid maps. The binding energy and inhibition constant for each pose were calculated and the best-selected poses were marked using the Discovery Studio Visualizer (24). The binding affinity and the interface are of the sACE2 protein with INS1 in the presence of nicotine was calculated again using PDBePISA.

**Results**

**Structural characteristics of the homology modelled ACE2 and the spike (S1) protein of SARS-CoV-2**

The downloaded ACE2 sequence was chosen for the modelling performed using SWISS-MODEL by which the crystal structure with PDB ID: 6m18 was enlightened as a potential template with an amino acid identity of 100% and query coverage of 99%. The modelled structure comprised of both extracellular and helical domains with the amino acids ranging from 21 to 768. Structure validation using the Ramachandran plot has confirmed 91.4% residues in the favoured region; 8.2% residues in the additional allowed region; 0.4% residues in the generously allowed region and 0% residues in the unfavourable region. From the modelled structure, the soluble ACE2 structure was obtained with the amino acid residues ranging from 21 to 708. For INS1 protein, 6VSB was the potential template with the structural homology of 99.17% and a query coverage of 95% (28). Domain details were procured from the literature study. The whole 3D structure model comprised of 27 to 1146 residues that got partitioned into S1 and S2 domain. Of the total 1146 residues, the S1 domain constitutes about 27-541 residues, while, the S2 domain constitutes about 778-1213 residues. Further, in-detail classification of the INS1 protein unravels the N-terminal domain (NTD) (aa 27-305); Receptor Binding Domain (aa 319-436 and aa 509-541); Receptor Binding Motif (aa 437-508); fusion peptide (aa 788-806); heptad repeat region ( aa 912-984); and the heptad repeat region 2 (aa 1163-1213). These structural conformations could not be modelled because of the inaccessibility of its template. As per the structure validation, Ramachandran Plot test results revealed 86.1% residues to exist within the most favoured region; 11.9% residues in the additional allowed region; 11.9%; the generously allowed region was 1.7% and the disallowed region was 0.3%.

**Structural characteristics of ACE2 binding with the spike (S1) protein of SARS-CoV-2**

ACE2 was docked with the INS1 wherein the RBD of INS1 was targeted against the ACE2 protein. Of the generated ten clusters, the one with the least HADDOCK score was considered for further examination. Three salt bridges were spotted with eleven hydrogen bonds with the ACE2-INS1 docking. Predominantly, the charged residues of ACE2 were involved in the interface interaction. A profound structural analysis was carried out in our previous study (29). The binding affinity of ACE2 with SARS-CoV-2 alone showed -15.7 kcal/mol with the measured interface area of 2057.3Å$^2$ between the ACE2 and SARS-CoV-2 RBD. While upon nicotine interaction, the binding affinity ACE2 with SARS-CoV-2 got drastically reduced to -12.8 kcal/mol with the constrained interface area of 1933.6 Å$^2$.

**Structural modulations inflicted by nicotine binding with the ACE2-SARS-CoV-2 protein complex**

With the help of localized docking approach performed using AutoDock software, nicotine was docked within the active site pocket of the modelled-soluble ACE2 protein (highly expressed in the arterio ventricular cells). Our docking results unveiled the nicotine's ability to bind profoundly into the pocket of ACE2 protein (Fig. 1a). It has been found that the positively charged His401 is indulged in direct contact with the nicotine. Also, the His401 has established the concomitant interactions with the Asp382, gly405, His378 and Tyr385 residues. All these residues were localized in the near distal regions of the protein-protein interface (Fig. 2a). After docking nicotine with the ACE2-SARS-CoV-2 complex, it has been found that nicotine establishes a stable interaction with Asp 368 that further interacts with the Thr362, Lys363, Thr365, Thr371 and Ala372 residues (Fig. 2b). All five amino acid residues engaged in these interactions were located at the proximal site to 353-357 segment of ACE2 protein that are actively participated in the stable interaction with the SARS-CoV-2 spike glycoprotein. The measured binding energy of nicotine with the soluble ACE2 was found to be -5.24 kcal/mol with the measured inhibition constant of 151.69 µM. Contradictorily, upon the nicotine's interaction with the ACE2-SARS-CoV-2 complex, its binding affinity got increased with -6.33 kcal/mol with the reduced inhibition constant of 22.95 µM. With the nicotine being binding to the active pocket site of sACE2, the SARS-CoV-2 binds to the spike protein binding site of ACE2. However, nicotine neither established any stable interaction nor developed an inhibitory contant value while interacting with sACE2 receptor alone that contradicts the results shown above with its interaction with the sACE2-SARS-CoV-2 complex. This in-silico observation has vividly explained that nicotine binding could produce pronounced effects on the ACE2 structure only while it gets sensitized with the SARS-CoV-2 binding, and not on the ACE2 structure alone. Indeed this would clearly define the noticeable benefits of nicotine exposure in the COVID-19 affected patients over the unaffected healthy individuals. Moreover, with this steady allosteric interaction of nicotine with ACE2 receptor could profusely intervene its binding with the spike (S1) protein of SARS-CoV-2 thereby showing its ability to tackle the COVID-19 virulence at a much lower dose on the COVID-19 patients.

**Discussion**

To manage the emerging COVID-19 virulence and its unprecedented transmission among the humans, the host-based ACE2 receptor has gathered immense attention among the clinicians (30). Since the COVID-19 acts as a potent agonist for ACE2 receptor (6), monoclonal antibodies targeting ACE2 has been proposed to be an effective treatment strategy for COVID-19 management (31). It has been shown that COVID-19 interaction with the ACE2 receptor would hinder its function causing severe respiratory ailments with the increased angiotensin II levels (6). Based on the recent clinical evidence that witnessed only mild adverse symptoms on the COVID-19 patients who are habituated to smoking than their counterparts (12, 32); we have planned to study the effect of nicotine binding with the in-silico modelled soluble ACE2 that is widely expressed on the arterial ventricular epithelial cells. This study is in no way tries to glorifiy the intake of nicotine through smoking. But it was mainly intended to understand the role of nicotine in inhibiting the SAR-COV2 protein from interacting with souble ACE2. Our docking study results have shown that nicotine has clearly established a fragile interaction with the naïve soluble ACE2 receptor. Nevertheless, under the same set-up, the nicotine has established a stable interaction with the ACE2 receptor that got sensitized in-prior with the COVID-19 binding. Also, nicotine's profound affirmative for ACE2-SARS-CoV-2 complex is highly likely to intervene in the interaction of ACE2 with the conserved amino acid residues of the spike (S1) protein of SARS-CoV-2 affecting its virulence. The study results also emphasize that even at a much lower concentration the nicotine would be able to hinder the interaction of SARS-CoV-2 with the ACE2 receptor. This might benefit the severely ill COVID-19 patients who are refrained from nicotine usage to overcome their adverse clinical symptoms by getting exposed to a much lower dose of nicotine that demands further clinical validation.

**Acknowledgement**

The authors would like to acknowledge technical support from the School of Biotechnology and Bioinformatics, D.Y. Patil Deemed to be University for providing access to the softwares for biological data analysis. This work was not funded by any external funding agencies.

**Competing interests**

None

**Figures**

Figure 1. sACE2 docked with Nicotine. The active site pocket of sACE2 was occupied by nicotine near the SARS-COV2 binding site.

Figure 2 sACE2 and sACE2-SARS-COV2 combo docked with nicotine (a) In sACE2 nicotine preferred to bind deep into the active site pocket near the SAR-COV2 binding site. But still they were quite away from the sACE2 protein SARS-COV2 binding site.(b) Nicotine docked with sACE2-SARS-COV2 docked protein which is like the disease state. Here, nicotine prefer to bind proximal to the spike protein binding site

**FIGURES**

**Figure 1**

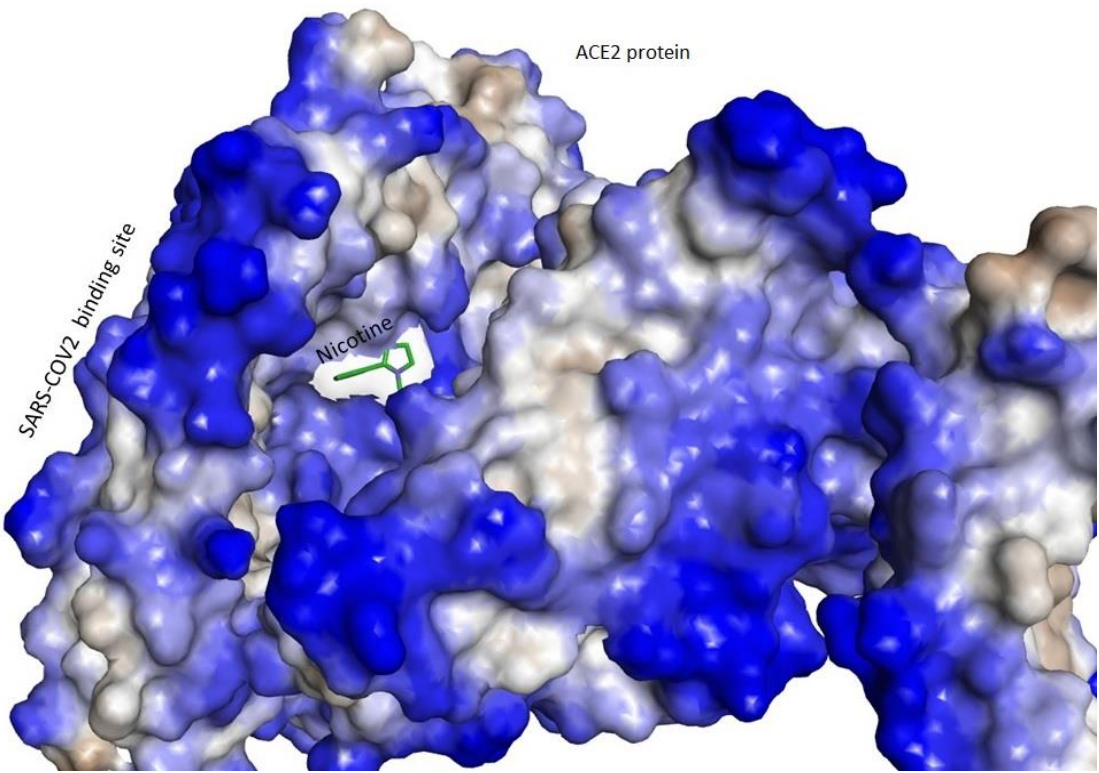

**Figure 2(a)**

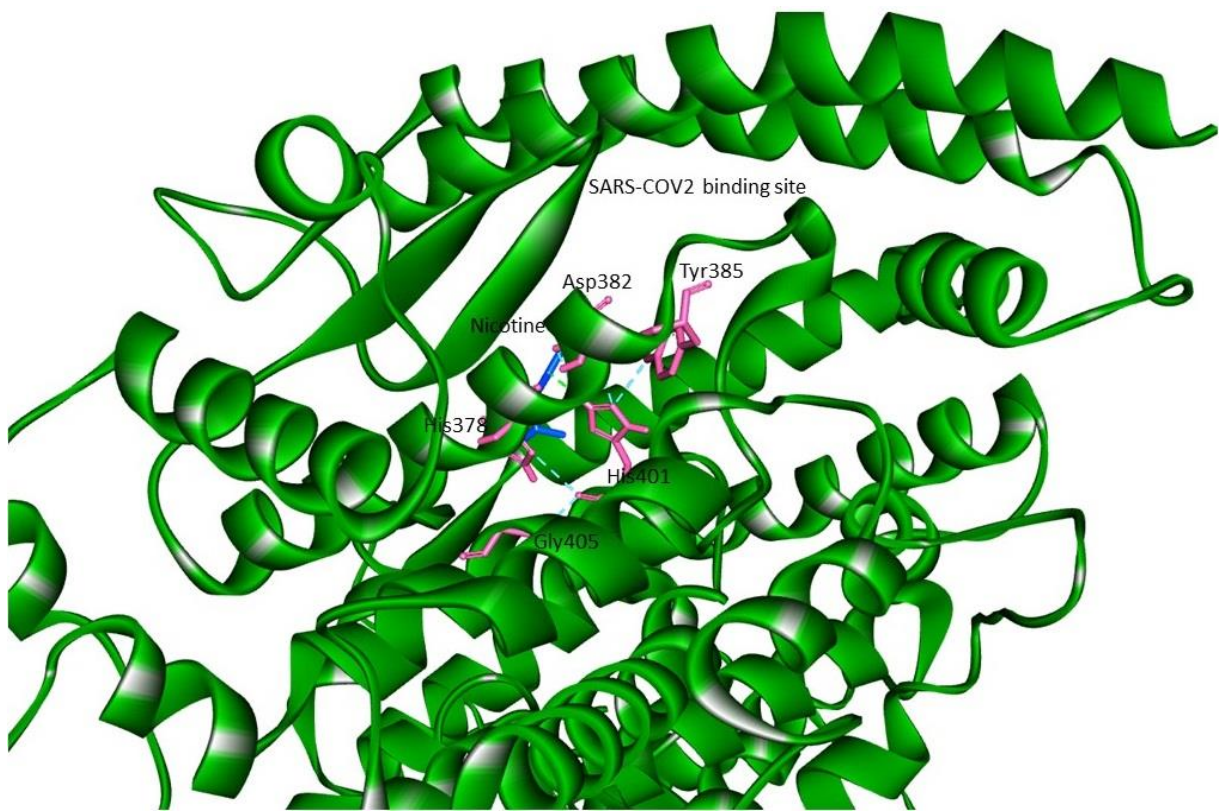

**Figure 2(b)**

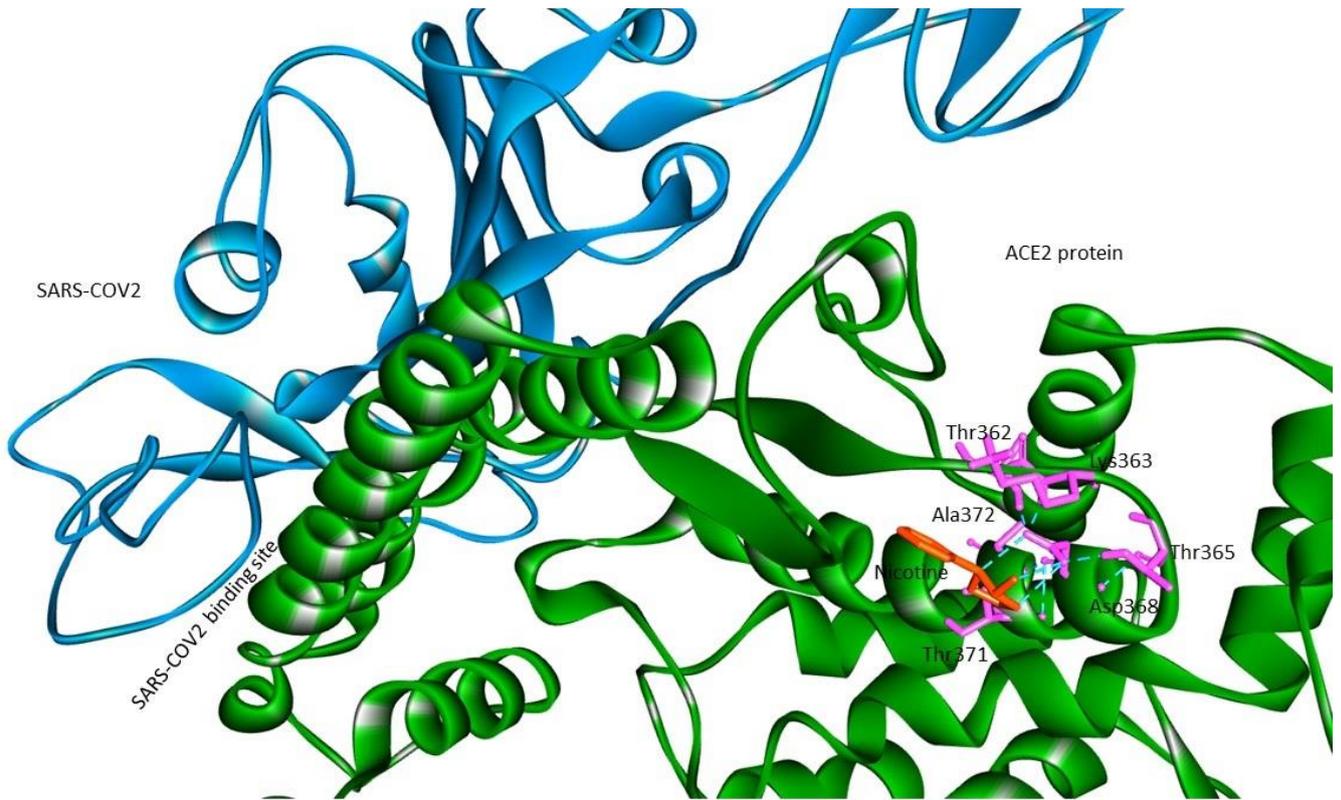